# *Role of Time-Reversal Symmetry in the Dynamical Response of "One-Way" Nonlinear Devices*

David E. Fernandes[1*], Mário G. Silveirinha[2†]

[1]University of Coimbra, Department of Electrical Engineering – Instituto de Telecomunicações, 3030-290 Coimbra, Portugal

[2]University of Lisbon, Instituto Superior Técnico and Instituto de Telecomunicações, Avenida Rovisco Pais, 1, 1049-001 Lisboa, Portugal

**Abstract**

We study the role of time-reversal symmetry on the dynamical response of nonlinear optical systems that behave as unidirectional ("one-way") devices. It is shown that lossless nonlinear materials, despite being nonreciprocal, are typically time-reversal invariant. This property raises an apparent paradox because time-reversal invariant systems are forcibly bi-directional. Here, we present a solution for this conundrum, and theoretically explain why the "one-way" behavior can indeed be compatible with the time-reversal invariance. It is found that in the time-reversed problem the incident waves have a variation in time that is generally incompatible with the adiabatic approximation. Due to this reason the adiabatic approximation fails to predict the bi-directional nature of nonlinear system. We discuss the implications of this finding in the performance of practical nonlinear "one-way" devices.

[*]E-mail: dfernandes@co.it.pt

[†]To whom correspondence should be addressed: E-mail: mario.silveirinha@co.it.pt

# I. Introduction

Reciprocity is a fundamental property of most photonic systems [1-4]. In a reciprocal system, the transmission level between any two given points is independent of the propagation direction, i.e., it remains invariant if the emitter and detector are interchanged. The reciprocal properties of electromagnetic systems are a consequence of linearity, conservation of energy, and time-reversal symmetry ($\mathcal{T}$) [4]. A link between reciprocity and time-reversal invariance in linear systems can be established even when the electromagnetic response of the system is dissipative [4].

Nonreciprocal devices – such as isolators, circulators or gyrators – are of fundamental importance in telecommunications systems [5-7]. There are different ways to obtain a nonreciprocal response, for example, by breaking the time-reversal symmetry [8-11], by exploiting nonlinearities [12-22], using systems with spacetime-modulations or with mechanical motion [23-27]. In reality the different mechanisms are not fully independent. For instance, the standard way to break the $\mathcal{T}$-symmetry is by applying a static magnetic bias to a material. This mechanism itself exploits the nonlinear response of charged particles to an applied **B**-field: the magnetic component of the Lorentz force depends both on the field and on the particle velocity. Thus, the nonreciprocity due to the magnetic bias essentially exploits a static-type nonlinearity, which drives the material to a new point of operation where the response to a dynamical (weak) signal is linear and nonreciprocal [28]. Similarly, the nonreciprocity of systems with active elements [29-31] or with an electric drift current bias [32-36] can also be understood in terms of static-type nonlinearities.

One important point which was little discussed in the recent literature is that even though nonlinear systems are nonreciprocal, the electrodynamics can remain time-reversal invariant if the dissipation is weak [17]. As is well-known, a fundamental consequence of the $\mathcal{T}$-symmetry is that the wave propagation is forcibly bi-directional [3-4]. On the other hand, nonlinear systems can have pronounced bi-stable transmission loops that can create the conditions for one-way propagation regimes [12-20]. The objective of this article is to address these seemingly incompatible properties and highlight some fundamental constraints on the dynamical response of nonlinear systems due to the time-reversal symmetry. Other limitations of nonlinear devices in optical isolation were discussed previously in the literature [37-39].

## II. Time-Reversal Symmetry in Nonlinear Devices

Here, we present a general argument that justifies why lossless nonlinear devices are typically $\mathcal{T}$-invariant. The analysis extends that of our previous work [17]. Some consequences of $\mathcal{T}$-symmetry in nonlinear acoustic systems were discussed in [40], while different aspects of the role of $\mathcal{T}$-symmetry in nonlinear optics were analyzed in [41, 42].

The Maxwell equations in a non-magnetic material system are:

$$\nabla \times \mathbf{H}(\mathbf{r},t) = \mathbf{j}(\mathbf{r},t) + \partial_t \mathbf{D}(\mathbf{r},t), \qquad \nabla \times \mathbf{E}(\mathbf{r},t) = -\mu_0 \partial_t \mathbf{H}(\mathbf{r},t), \qquad (1)$$

with $\partial_t = \partial/\partial t$, $\mathbf{D}(\mathbf{r},t) = \varepsilon_0 \mathbf{E}(\mathbf{r},t) + \mathbf{P}(\mathbf{r},t)$ is the electric displacement vector and $\mathbf{P}$ is the polarization vector. Without loss of generality, let us suppose that $\mathbf{P}(\mathbf{r},t)$ and $\mathbf{E}(\mathbf{r},t)$ are linked by a nonlinear Lorentz-type dispersive model, such that:

$$\partial_t^2 \mathbf{P}(\mathbf{r},t) + \omega_0^2 \mathbf{P}(\mathbf{r},t) = \varepsilon_0 \omega_0^2 \overline{\chi}(\mathbf{r}, \mathbf{E}(\mathbf{r},t)) \cdot \mathbf{E}(\mathbf{r},t), \qquad (2)$$

where the susceptibility tensor $\bar{\bar{\chi}}$ generically depends on the instantaneous electric field intensity and on the space coordinates. Here, $\omega_0$ plays the role of a resonance frequency. For $\omega_0 \to \infty$ the above equation reduces to the dispersionless model $\mathbf{P}(\mathbf{r},t) = \varepsilon_0 \bar{\bar{\chi}}(\mathbf{r}, \mathbf{E}(\mathbf{r},t)) \cdot \mathbf{E}(\mathbf{r},t)$. It is straightforward to verify that if some dynamical fields $\mathbf{E}(\mathbf{r},t), \mathbf{H}(\mathbf{r},t), \mathbf{P}(\mathbf{r},t), \mathbf{j}(\mathbf{r},t)$ satisfy Eqs. (1)-(2) then the time reversed fields $\mathbf{E}^{\text{TR}}(\mathbf{r},t), \mathbf{H}^{\text{TR}}(\mathbf{r},t), \mathbf{P}^{\text{TR}}(\mathbf{r},t), \mathbf{j}^{\text{TR}}(\mathbf{r},t)$ defined by,

$$\mathbf{E}^{\text{TR}}(\mathbf{r},t) = \mathbf{E}(\mathbf{r},-t), \qquad \mathbf{P}^{\text{TR}}(\mathbf{r},t) = \mathbf{P}(\mathbf{r},-t), \tag{3a}$$

$$\mathbf{H}^{\text{TR}}(\mathbf{r},t) = -\mathbf{H}(\mathbf{r},-t), \qquad \mathbf{j}^{\text{TR}}(\mathbf{r},t) = -\mathbf{j}(\mathbf{r},-t), \tag{3b}$$

also do. Thus, the nonlinear Maxwell's equations (1)-(2) are invariant to a time reversal.

The previous analysis can be extended to a wide range of nonlinear couplings. For instance, a system with multiple resonances may be modeled by $\mathbf{P}(\mathbf{r},t) = \sum_i \mathbf{P}_i(\mathbf{r},t)$, with $\partial_t^2 \mathbf{P}_i(\mathbf{r},t) + \omega_{0,i}^2 \mathbf{P}_i(\mathbf{r},t) = \varepsilon_0 \omega_{0,i}^2 \bar{\bar{\chi}}_i(\mathbf{r}, \mathbf{E}(\mathbf{r},t)) \cdot \mathbf{E}(\mathbf{r},t)$, also gives a time-reversal invariant response. Furthermore, any nonlinear conservative material response that is described by a time-reversal invariant (non-quadratic) Lagrangian density yields time-reversal invariant nonlinear dynamics. For instance, suppose that the Lagrangian density is:

$$\mathcal{L} = \frac{1}{2} \sum_i \frac{1}{\omega_{p,i}^2} \left( \partial_t \mathbf{P}_i \cdot \partial_t \mathbf{P}_i - \omega_{0,i}^2 \mathbf{P}_i \cdot \mathbf{P}_i \right) + \mathcal{L}_{\text{int}}(\mathbf{P}_1, \mathbf{P}_2, ...) + \varepsilon_0 \mathbf{E} \cdot \sum_i \mathbf{P}_i. \tag{4}$$

The Lagrangian density is evidently time-reversal invariant. The Euler-Lagrange equations $\dfrac{d}{d}\dfrac{\partial \mathcal{L}}{\partial(\partial_t \mathbf{P}_i)} = \dfrac{\partial \mathcal{L}}{\partial \mathbf{P}_i}$ give:

$$\partial_t^2 \mathbf{P}_i + \omega_{0,i}^2 \mathbf{P}_i - \omega_{p,i}^2 \frac{\partial \mathcal{L}_{\text{int}}}{\partial \mathbf{P}_i} = \varepsilon_0 \omega_{p,i}^2 \mathbf{E}, \tag{5}$$

which describes a time-reversal invariant nonlinear system when the interaction between the different oscillators $\mathcal{L}_{int}(\mathbf{P}_1,\mathbf{P}_2,...)$ does not reduce to a quadratic form.

Generically, most non-dissipative nonlinear couplings yield $\mathcal{T}$-invariant dynamics [43-45]. For example, an instantaneous $\chi^{(3)}$ type coupling ($\mathbf{P}(\mathbf{r},t) = \varepsilon_0 \left(\chi^{(3)}\mathbf{E}(\mathbf{r},t)\cdot\mathbf{E}(\mathbf{r},t)\right)\cdot\mathbf{E}(\mathbf{r},t)$) is also $\mathcal{T}$-invariant. In contrast, a system with a constitutive relation of the type $\partial_t^2\mathbf{P}(\mathbf{r},t) + \Gamma\partial_t\mathbf{P}(\mathbf{r},t) + \omega_0^2\mathbf{P}(\mathbf{r},t) = \varepsilon_0\omega_0^2\overline{\chi}(\mathbf{r},\mathbf{E}(\mathbf{r},t))\cdot\mathbf{E}(\mathbf{r},t)$, with $\Gamma$ a collision frequency that models the loss in the system, yields a response that is not $\mathcal{T}$-invariant. This happens due to the dissipative term $\Gamma\partial_t\mathbf{P}(\mathbf{r},t)$ [46].

A consequence of the $\mathcal{T}$-symmetry is that if some time dynamics is physically allowed in the system, then the time-reversed dynamics may also be observed if the initial conditions are suitably chosen [3, 47]. In other words, in a $\mathcal{T}$-invariant system any propagation process is ultimately reversible. This property has been exploited in systems with a linear response to focus electromagnetic waves on subwavelength spots [48-52]. If some electromagnetic pulse can propagate, let us say from left to right, through some nonlinear non-dissipative $\mathcal{T}$-invariant channel or device with negligible reflections, then the time-reversed pulse will follow the time-reversed trajectory, from right to left, also with negligible reflections [3, 20]. Thereby, $\mathcal{T}$-invariant nonlinear devices with negligible dissipation are fundamentally bi-directional.

Interestingly, many recent works have demonstrated a one-way propagation through two-port nonlinear devices such that $T_{1\rightarrow 2} \neq 0$ and $T_{2\rightarrow 1} \approx 0$, with $T_{i\rightarrow j}$ the transmissivity from port $i$ to port *j* [12-20]. The asymmetric response of a nonlinear device to both directions of incidence is typically rooted in pronounced bi-stable transmission loops that make the system alternate between "transparent" and "opaque" states depending on the excitation port. A device with $|T_{1\to 2}| \gg |T_{2\to 1}| \approx 0$ is usually referred to as an "electromagnetic diode". A common figure of merit for the electromagnetic diode is the value of $T_{1\to 2}/T_{2\to 1}$, which for a fixed input power should be much larger than one, with $T_{1\to 2}$ desirably near unity. Furthermore, by using coupled nonlinear resonances it may be possible to design broadband nonlinear systems with negligible absorption such that $T_{1\to 2} \approx 1$ and $T_{2\to 1} \approx 0$ [16, 20]. This property is apparently at odds with the time-reversal invariance of the nonlinear system, which as discussed previously forcibly implies bi-directional propagation. In what follows, we resolve this conundrum, and explain the reason why the electromagnetic diode operation can be compatible with the time-reversal invariance.

### III. Electromagnetic Diode

#### A. *Physical model*

To illustrate the key theoretical ideas, we choose as an example a nonlinear device that consists of a two-sided mushroom metamaterial slab (see Fig. 1a). The mushroom metamaterial is formed by two arrays of metallic patches, with each pair of patches connected through a metallic wire [53-57] and a nonlinear capacitor (varactor). In our setup, there is an offset between the center of the patches and the wires connection. This offset is required so that a normally incident plane wave can induce a current on the metallic wires. The nonlinear mushroom structure is excited by plane waves that propagate along +*z* and/or –*z*-directions and have the electric field oriented along the +*x*-direction. The nonlinear mushroom metamaterial may be used to absorb high-power signals [58, 59] or as an electromagnetic switch controlled by the intensity of the incident wave [17, 60].

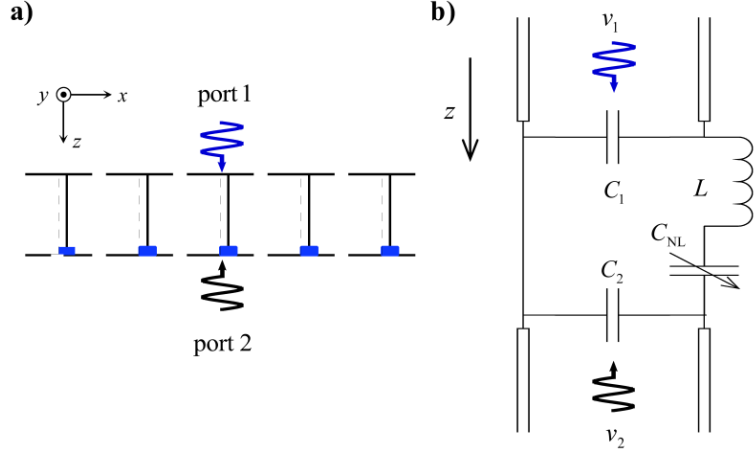

**Fig. 1. a)** Side view of the geometry of the two-sided mushroom slab. The wires are embedded in air and are connected to metallic patches through lumped nonlinear capacitors at the bottom interface, represented in the figure as blue rectangles, and through ideal short-circuits at the top interface. **b)** Equivalent circuit model consisting of two transmission lines connected to a nonlinear circuit formed by lumped elements. The circuit consists of two shunt capacitors $C_1$ and $C_2$, and of an inductor $L$ and a nonlinear capacitor $C_{NL}$ connected in series.

We use the equivalent transmission line circuit of Fig. 1b to model the nonlinear time dynamics of the metamaterial. The transmission lines have characteristic impedance $Z_0 = \sqrt{\mu_0/\varepsilon_0}$, with $\mu_0, \varepsilon_0$ the vacuum permeability and permittivity, respectively, and are connected to two shunt capacitors $C_1$ and $C_2$ that describe the effective response of the patch arrays [56, 61], and to the series of an inductance $L$ that models the metallic wires and a nonlinear capacitor $C_{NL}$ that determines the response of the varactor. In the model, we use $C_1 \neq C_2$ to account for the structural asymmetry created by the nonlinear element. Without loss of generality, it is supposed that the capacitance of the nonlinear capacitor depends on the charge $q$ as

$C_{NL}(q) = C_0 \left( 1 + \dfrac{\alpha_{NL}|q|^M}{1 + \beta_{NL}|q|^M} \right)$. The capacitance of the nonlinear element saturates for a large $q$,

consistent with the typical physical response of most commercially available varactors. Throughout the article, it is assumed that $C_1 = 0.22\text{pF}$, $C_2 = 0.26\text{pF}$ and that $L = 3.5\text{nH}$. These parameters were chosen to mimic the response of a particular metamaterial structure near the operating frequency $f = 14.56\text{GHz}$ (see Appendix A). The nonlinear varactor is modeled by $C_0 = 40\text{fF}$, $\alpha_{\text{NL}} = 2.2 \times 10^{-17}/q_e^M$, $\beta_{\text{NL}} = 4\alpha_{\text{NL}}$, and $M = 2.5$, with $q_e = 1.6 \times 10^{-19}\text{C}$ the elementary charge.

The voltages in the transmission lines can be decomposed as $v_1 = v_1^+ + v_1^-$ and $v_2 = v_2^+ + v_2^-$, with the "+" ("−") superscript associated with the waves that propagate along the $+z$ ($-z$) direction. Similarly, the currents in the transmission line are $i_n = i_n^+ - i_n^- = (v_n^+ - v_n^-)/Z_0$, with $n = 1, 2$. The voltages at ports 1 and 2 can be written in terms of the incident waves as:

$$v_1 = 2v_1^+ - Z_0 i_1, \tag{6a}$$

$$v_2 = 2v_2^- + Z_0 i_2. \tag{6b}$$

From Fig. 1b, we see that $v_1 = v_{\text{load}} + v_2$, where $v_{\text{load}} = L\, di_L/dt + v_{\text{NL}}$ is the voltage across the inductor and the nonlinear capacitor. The voltage $v_{\text{NL}}$ is defined as $v_{\text{NL}} = q/C_{\text{NL}}$. Using $i_L = dq/dt$ it follows that:

$$v_1 = L\frac{d^2 q}{dt^2} + \frac{q}{C_{\text{NL}}} + v_2 \tag{7}$$

The charge conservation in each transmission line implies that $i_L = i_1 - C_1 \dfrac{dv_1}{dt} = i_2 + C_2 \dfrac{dv_2}{dt}$. Therefore,

$$i_1 - i_2 = C_2 \frac{dv_2}{dt} + C_1 \frac{dv_1}{dt} \tag{8}$$

By combining Eqs. (6)-(8), we obtain a set of master equations that determine the dynamics of $v_1, v_2, q$ as a function of the incident waves $v_1^+, v_2^-$:

$$L\frac{d^2q}{dt^2} + \frac{q}{C_{NL}} + (v_2 - v_1) = 0, \qquad (9a)$$

$$Z_0 C_1 \frac{dv_1}{dt} + v_1 = 2v_1^+ - Z_0 \frac{dq}{dt}, \qquad (9b)$$

$$Z_0 C_2 \frac{dv_2}{dt} + v_2 = 2v_2^- + Z_0 \frac{dq}{dt}. \qquad (9c)$$

Note that the capacitance of the nonlinear element $C_{NL}$ is a function of $|q|$.

### B. Steady-state response

For simplicity, we use a rotating wave approximation such that a generic voltage is of the form $v(t) = V(t) e^{-i\omega t}$, with $V(t)$ the amplitude "envelope" and $\omega$ the oscillation frequency ($\omega/2\pi = 14.56$ GHz). When the envelope of the incident wave $V^{inc}$ is independent of time, the system reaches a stationary state such that the envelopes of all voltages are also time independent ($v(t) = V_\omega e^{-i\omega t}$). The steady-state solutions are found by solving Eq. (9) with $\partial/\partial t = -i\omega$.

Figure 2 shows the calculated steady-state envelope amplitude of the charge stored in the nonlinear element as a function of $|V^{inc}|$ for individual excitations of ports 1 (blue line, $|V^{inc}| = |V_1^+|$) and 2 (black line, $|V^{inc}| = |V_2^-|$). Interestingly, for some values of $|V^{inc}|$ there are up to three allowed operation points. The actual point of operation can only be determined by knowing the precise time evolution of the system before the steady state is reached. This conditional choice of the point of operation can force abrupt jumps in the physical response when the amplitude of the envelope of the incident wave is varied [17, 60].

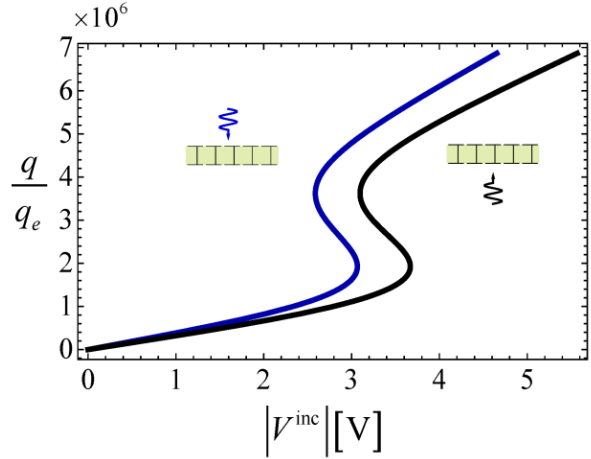

**Fig. 2.** Charge stored at the nonlinear element as a function of the incident wave amplitude calculated in stationary state for $\omega/2\pi = 14.56 \text{GHz}$.

## C. "Electromagnetic Diode" operation

The relation between $|V^{inc}|$ and the charge $|q|$ is not univocal. As is well-known, a bi-stable response of the system gives rise to hysteresis loops in the transmission properties of the system [17, 60]. Figure 3 shows the transmissivity ($|T|^2$) curves calculated as a function of the amplitude of the incident wave for individual port excitations. The transmission coefficient is defined as $T_{top} = V_2^+/V_1^+$ for a port 1 (top) excitation and $T_{bot} = V_1^-/V_2^-$ for a port 2 (bottom) excitation. As seen, the transmission level depends on the point of operation. Possible jumps (phase transitions) between the branches of the hysteresis loops are highlighted with the vertical arrows. The transitions occur when the amplitude of the incident wave is increased (for the upward arrow) or decreased (for the downward arrow) beyond some critical value. Depending on the time dynamics, the metamaterial can switch abruptly from a transparent state ($|T|^2 \approx 1$) to an opaque state ($|T|^2 \approx 0$) or vice-versa. The response of the metamaterial slab can be highly asymmetric for an excitation with intensity on the order of $2.7 < |V^{inc}| < 3.2 \text{ [V]}$. In this regime, the system behaves as a nearly ideal "electromagnetic diode", such that $|T_1|/|T_2| \gg 1$ with $|T_1|^2 \approx 1$ and $|T_2|^2 \approx 0$, i.e. it is almost transparent to a port 1 excitation and approximately opaque to a port 2 excitation [17].

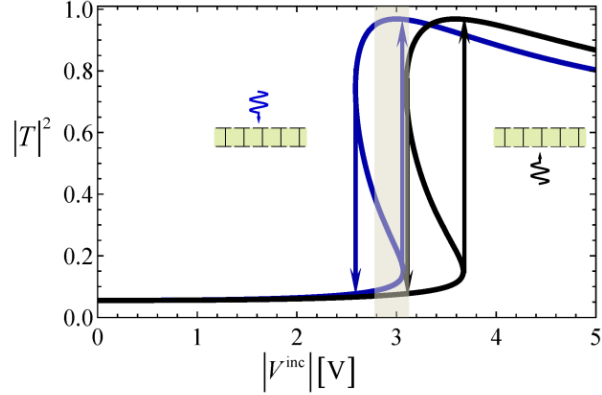

**Fig. 3.** Transmissivity of the metamaterial slab as function of the amplitude of the incident wave calculated for port 1 (blue curve) and port 2 (black curve) excitations. The vertical arrows represent the jumps of the bi-stable hysteresis cycle occurring when the amplitude of the excitation is either decreasing or increasing. The oscillation frequency is $\omega/2\pi = 14.56\text{GHz}$.

Typically, the dynamics of nonlinear devices is predicted using an adiabatic approximation that assumes that the system state follows the hysteresis loop obtained for the steady-state excitation. Such an approximation can be rather accurate when the envelope of the incident waves varies slowly in time. To illustrate this property, we consider a slowly time-varying incident pulse (blue line in Fig. 4a) with a peak amplitude $|V^{inc}| \approx 3.1\text{V}$ that gradually decreases to $|V^{inc}| \approx 2.8\text{V}$ (the interval $2.8\text{V} < |V^{inc}| < 3.1\text{V}$ corresponds to the shaded region in Fig. 3). The format of the pulse is chosen so that port 1 is operated in the bistable regime and in the bright state for most of the duration of the pulse: the peak amplitude of the pulse lies just outside the hysteresis loop and the small time-decaying slope ensures that the point of operation slides along the top branch of the hysteresis loop. On the other hand, for a port 2 excitation with this pulse, the system remains in the dark state (bottom branch of the hysteresis loops) during the entire excitation. Thereby, the adiabatic approximation predicts that in the time interval where $2.8\text{V} < |V^{inc}| < 3.1\text{V}$ the system will be operated in a nearly transparent state (top branch of the blue colored hysteresis loop) for a port 1 excitation, and in a nearly opaque state (bottom branch of the black colored hysteresis loop) for a port 2 excitation. The detailed time dynamics obtained with the adiabatic approximation is shown in Figs. 4a-b. In the figure we use the notations $v_1^{inc} = v_1^+$, $v_2^{inc} = v_2^-$ for the incident waves and $v_1^{sct} = v_1^-$, $v_2^{sct} = v_2^+$ for the scattered waves.

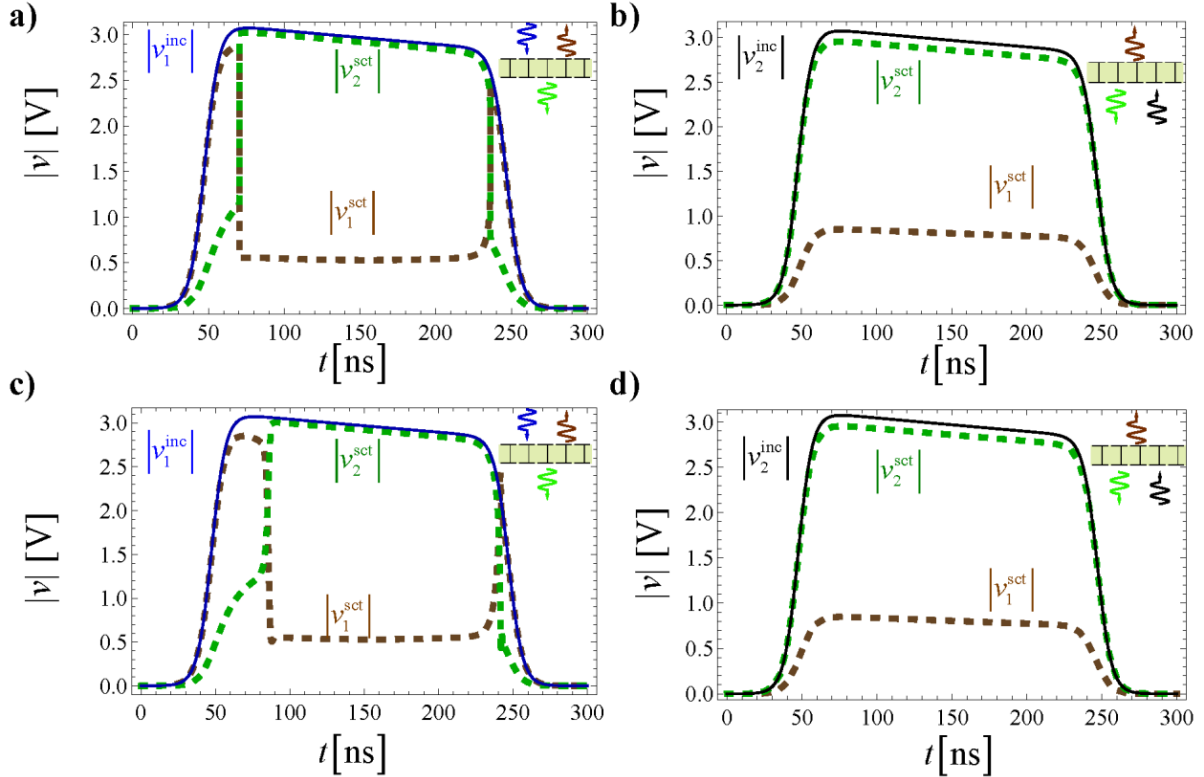

**Fig. 4. a)** Envelopes of the incident wave $v_1^{inc}$ (blue solid curve) at port 1 and of the scattered signals $v_1^{sct}$ (brown dashed curve) and $v_2^{sct}$ (green dashed curve) as a function of time calculated using an adiabatic approximation. **b)** Similar to **a)** but for a wave ($v_2^{inc}$) incident at port 2. **c)** and **d)** Similar to **a)** and **b)** but calculated using the exact model.

The exact time dynamics of the system was calculated by numerically solving Eq. (9). As seen in Figs. 4c-d, the exact result concurs generically rather well with the adiabatic approximation for both excitations, confirming the validity of the adiabatic approach. The fingerprint of the phase transitions for a port 1 excitation can be seen in the beginning and in the end of the pulses depicted Figs. 4a and 4c. The phase transitions are accompanied by abrupt variations (discontinuous in the adiabatic approximation) of the scattered wave envelopes which cause undesired reflections. This is better visualized in Fig. 5, which depicts the scattered signals near the time instants where the phase transitions take place.

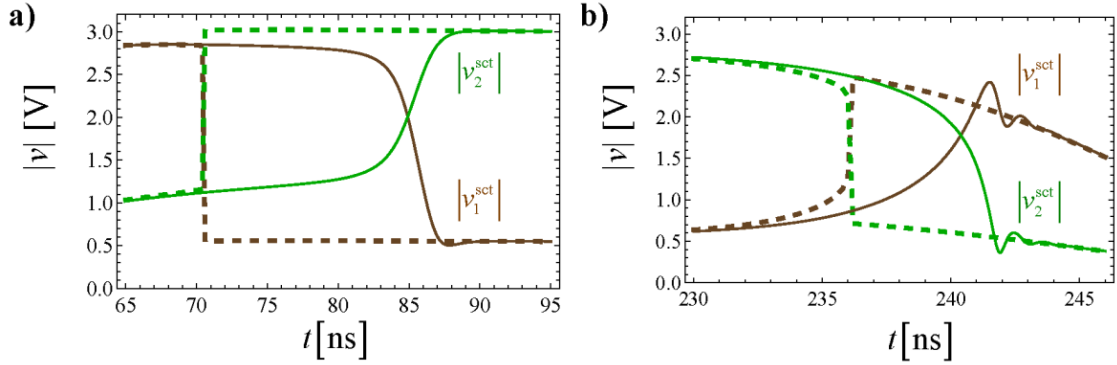

**Fig. 5. a)** Envelopes of the scattered signals for a port 1 excitation calculated at port 1 $v_1^{sct}$ (brown curve) and at port 2 $v_2^{sct}$ (green curve) as a function of time near the first phase transition obtained using the adiabatic approximation (dashed curves) and the exact model (solid curves). **b)** Similar to **a)** but calculated near the second phase transition.

Near these time instants there is an evident discrepancy of the results predicted by the two calculation methods. The phase transitions limit the performance of the metamaterial slab as an electromagnetic isolator, as they may generate a narrow reflected pulse that propagates back to the source (see in particular the second phase transition in Figs. 4c and 5b). The emergence of these narrow-pulses is unavoidable as the phase transitions are associated with very high frequencies which effectively short-circuit the capacitors and open-circuit the inductors. Since for a port 2 excitation the pulse does not generate phase transitions, there are no high-frequency signals in that case. Qualitatively similar results can be obtained for other formats of the pulse with a similar peak amplitude and relatively slow rise and fall times. An example of such an excitation is a Gaussian pulse with an envelope $V(t) = V_0 e^{-\frac{(t-T_0)^2}{\Delta T^2}}$ with $\Delta T \omega / 2\pi > 2000$, $T_0 > 2\Delta T$ and $V_0 = 3.2\text{V}$ (not shown).

## IV. Time-Reversal

Consistent with previous works [12-20], the analysis of the previous section confirms that in steady-state a nonlinear device can imitate an electromagnetic diode for the individual excitation of its ports with the same pulse. Our system is lossless and thereby, as shown in Sect. II, $\mathcal{T}$-invariant. Indeed, it can be readily checked that if $v(z,t)$ and $i(z,t)$ are solutions of some scattering problem defined for the system of Fig. 1b, then the time-reversed solutions $v_{TR}(z,t) = v^*(z,-t)$ and $i_{TR}(z,t) = -i^*(z,-t)$ are also compatible with the wave propagation in the same system [4]. Since the time-reversal invariance implies that the system is bi-directional, how can this property be reconciled with the "electromagnetic diode" operation?

To begin with, we numerically verify the $\mathcal{T}$-invariance property of the metamaterial slab. Under a time-reversal operation the Poynting vector is flipped, and thereby the roles of the incident and scattered waves are exchanged. Thus, it follows that:

$$v_{i,\text{TR}}^{\pm}(t) = \left[ v_i^{\mp}(-t) \right]^*, \qquad (i=1,2). \qquad (10)$$

We numerically solved Eq. (9) using as the incident waves $v_{i,\text{TR}}^{\text{inc}}(t) = \left[ v_i^{\text{sct}}(-t) \right]^*$, where $v_i^{\text{sct}}(t)$ are the scattered waves obtained from the "direct" problems considered in Fig. 4c-d. Thus, ports 1 and ports 2 are simultaneously illuminated by the (time-reversed) waves that are scattered in the "direct problem". We will refer to this new problem as the "time-reversed (TR) problem". If the system is $\mathcal{T}$-invariant, all the energy must be returned to the port from which it originally came from in the direct problem. In other words, the numerical solution must be such that $v_{i,\text{TR}}^{\text{sct}}(t) = \left[ v_i^{\text{inc}}(-t) \right]^*$.

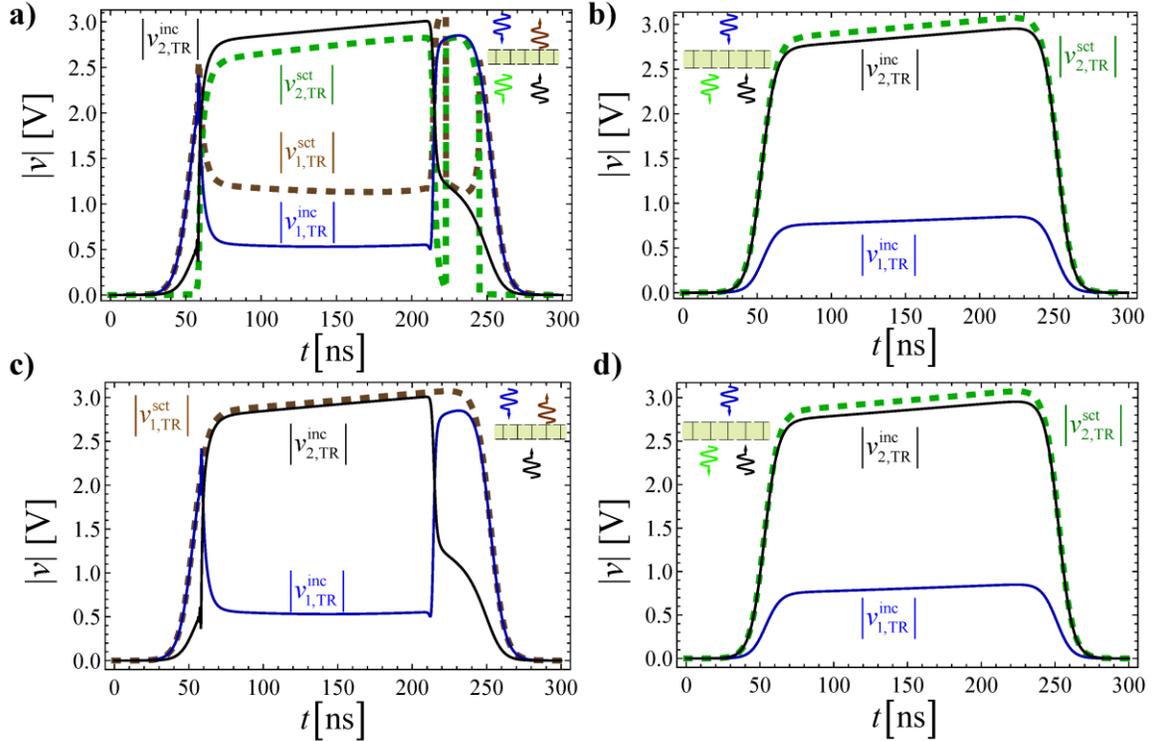

**Fig. 6. a)** Envelopes of the incident waves $v_{1,\text{TR}}^{\text{inc}}$ (blue curve) and $v_{2,\text{TR}}^{\text{inc}}$ (black curve) obtained from a time-reversal of the direct problem for a port 1 excitation (Fig. 4c) as a function of time. The scattered voltages calculated using an adiabatic approximation are $v_{1,\text{TR}}^{\text{sct}}$ (brown dashed curve) and $v_{2,\text{TR}}^{\text{sct}}$ (green dashed curve). **b)** Similar to **a)** but for incident waves obtained from a time-reversal of the direct problem for a port 2 excitation (Fig. 4d). The scattered signal calculated using an adiabatic approximation at port 2 is $v_{2,\text{TR}}^{\text{sct}}$ (green dashed curve). The scattered wave at port 1 vanishes ($v_{1,\text{TR}}^{\text{sct}} = 0$). **c)** Similar to **a)** but calculated using the exact model. The scattered wave at port 2 vanishes ($v_{2,\text{TR}}^{\text{sct}} = 0$) **d)** Similar to **b)** but calculated using the exact model.

Let us first consider the direct problem corresponding to the individual excitation of port 1 (Fig. 4c). For the corresponding TR problem, the nonlinear metamaterial is simultaneously illuminated with the strong signal $v_2^{sct,*}(-t)$ in port 2 and by the weak signal $v_1^{sct,*}(-t)$ in port 1. Thus, this TR problem is rather similar to an individual excitation of port 2, for which as seen from Fig. 4d, the device supposedly blocks the wave transmission. Accordingly, Fig. 6a shows that the solution of the TR problem calculated with the adiabatic approximation produces very strong reflections in port 2 during the time interval $60\,\text{ns} < t < 210\,\text{ns}$. In the TR problem the time origin is shifted by about 300ns so that all the incident waves have most of their energy in the range $t > 0$.

Importantly, in the TR problem, there is a dramatic deviation between the exact numerical solution of Eq. (9) and the prediction of the adiabatic approximation. As shown in Fig. 6c, the numerical solution of Eq. (9) is fully consistent with the time-reversal property and yields exactly $v_{i,TR}^{sct}(t) = \left[v_i^{inc}(-t)\right]^*$, such that all energy is returned to port 1 in the TR problem and the reflections in port 2 are precisely zero ($v_{2,TR}^{sct}(t) = 0$). The scattered wave in port 1 is $v_{1,TR}^{sct}(t) = \left[v_1^{inc}(-t)\right]^*$, which is the original incident wave in the direct problem but reversed in time and propagating back towards the generator.

The reason for the discrepancy between the adiabatic approximation and the exact solution is easy to uncover. It is rooted in the fact that the envelopes of the incident waves in the TR problem do not vary slowly in time near the instants $t \approx 60\,\text{ns}$ and $t \approx 210\,\text{ns}$. These fast variations are inherited from the phase transitions of the direct problem. Understandably, the adiabatic approximation fails to predict the dramatic impact of such fast signal variations, which occur in a rather short period of time as compared to the rise and fall times of the incident pulse envelope. Specifically, phase transition of the direct problem takes around 10 periods of oscillation ($10 \times 2\pi/\omega$). Clearly, the "electromagnetic diode" behavior can be observed only when the envelopes of the excitation signals vary slowly in time. If the envelopes of the excitation signals have high-frequency components – as in the time-reversed problem – the adiabatic approximation is inapplicable, and the bi-directional nature of the system can be revealed. The previous discussion unveils that the reason why the electromagnetic diode regime is compatible with the time-reversal symmetry and links it with the breakdown of the adiabatic approximation near the time instants associated with phase transitions.

We also solved the TR problem corresponding to a port 2 excitation (direct problem of Fig. 4d). In this case, the dynamical and the adiabatic approximation solutions agree rather well (Figs. 6b and 6d). In fact, in this case the excitation in the direct problem (Fig. 4d) does not generate phase transitions. Thus, the envelopes of the incoming waves in the TR problem also vary slowly with time.

We numerically checked (not shown) that for weakly dissipative varactors (e.g., when the loss is modelled by a series resistor on the order of $R = 0.2\Omega$) the response of the system remains quasi-time reversal invariant such that the results of Fig. 6c-d do not change significantly. If the loss level is increased beyond some threshold the quasi time-reversal invariance is no longer observed

It is important to underline that the discussed issues are of practical importance. Indeed, even when the envelope of the incident field varies slowly in time, the transmitted scattered signals can have high-frequency components due to the phase transitions suffered during the excitation period. In particular, the phase transitions can generate high-frequency pulses that are sent back to the generator, as illustrated in Fig. 4a. Fortunately, there is a relatively simple way of fixing this type of problems, as in principle the high-frequency signals generated by the phase transitions can be eliminated by suitably designed filters that absorb the high-frequency part of the scattered signals.

## V. Conclusions

In this article, we investigated the dynamical constraints arising due to the time-reversal symmetry in the operation of nonlinear "one-way" devices. It was discussed that lossless nonlinear systems are most often time-reversal invariant, and thereby are bi-directional. This property is seemingly at odds with the "one-way" diode-type operation reported in many works in the recent literature. Our analysis shows that the solution of the apparent paradox is related to the fact that "one-way" nonlinear devices inevitably suffer phase transitions that generate scattered signals with high-frequency components. If the scattered signals are sent back to the nonlinear device, including the high-frequency components, the bi-directional nature of the nonlinear device is inevitably uncovered. In practice, this non-ideal behavior can be avoided with filters that absorb the generated high-frequency components. We used a nonlinear time domain model to illustrate the main ideas and the relevant physical mechanisms. In particular, our theory highlights some limitations of the standard adiabatic approximation in dynamical problems that generate phase transitions and high-frequency fields.

## Appendix A: The mushroom metamaterial

The two-sided mushroom metamaterial consists of an array of metallic wires arranged in a square lattice [54-56] and connected to metallic patches [55, 57]. In this work we consider a lattice period $a = 1\,\text{cm}$ and wires with radius $r_w = 0.025a$. The wires are assumed to stand in air and are misaligned with respect to the geometrical center of the patches by a distance $d = a/4$ (highlighted in Fig. 1a as dashed curves). The thickness of the slab is $h = 4a$ and the separation between adjacent patches is $g = a/20$. At the top interface, the wires are connected to the metallic patches through an ideal short-circuit. At the bottom interface, the vias-to-patch connection is done using a nonlinear load (blue rectangles in Fig. 1a). The nonlinear load is modeled as a parallel plate-type capacitor filled with a nonlinear Kerr-type dielectric. The capacitance of the nonlinear element may be written as a function of the voltage $V_L$ across its terminals [60] as $C_{\text{NL}}(V_L) = \frac{\varepsilon_0 \varepsilon_{\text{cap}}^0 l_{\text{cap}}^2}{t_{\text{cap}}}\left(1 + \frac{\alpha}{t_{\text{cap}}^2}|V_L|^2\right)$, where $l_{\text{cap}}^2 = 9g^2$ corresponds to the capacitor cross-sectional area, $t_{\text{cap}} = 2g$ to its thickness, $\varepsilon_{\text{cap}}^0 = 2$ and $\alpha = \frac{3}{4}\frac{\chi^{(3)}}{\varepsilon_{\text{cap}}^0}$, with the third-order electric susceptibility taken equal to $\chi^{(3)} = 0.9 \times 10^{-9}\,\text{m}^2\text{V}^{-2}$.

In Ref. [17, 60] we developed a homogenization model to characterize the nonlinear response of the metamaterial which only considers the nonlinear self-action in the capacitor. This model is too complex to calculate the full nonlinear time dynamics of the metamaterial. To circumvent this limitation, we developed a simpler transmission-type model of the metamaterial (see Fig. 1b). In the model, the two shunt capacitors $C_1$ and $C_2$ describe the effective response of the patch arrays [56, 61]. The inductance $L$ models the wire array. The nonlinear capacitor $C_{\text{NL}}$ determines the response of the parallel plate-type capacitor filled with a nonlinear Kerr-type dielectric. The structural asymmetry due to the misalignment of the wires and patches is taken into account in the equivalent circuit model by considering $C_1 \neq C_2$.

The parameters of the transmission line model are tuned so that the steady-state responses predicted by the homogenization model of Ref. [17, 60] and the transmission line model agree nearby the oscillation frequency $f = 14.56\,\text{GHz}$. This frequency is associated with a Fano-type resonance [20, 60], which is exploited to enhance the sensitivity of the system to the nonlinear effects.

The top patch grid is modeled by $C_1 = 1.5 Y_g/(-i\omega)$, with $Y_g = -i\omega(\varepsilon_h + \varepsilon_0)(a/\pi)\log[\csc(\pi g/2a)]$ the patch grid admittance [61]. The bottom patch grid equivalent capacitance is $C_2 = 1.2 C_1$. The wire-medium slab response is modeled with the inductance $L = 1.113 \mu_0 \frac{a^2}{h}$. The nonlinear capacitor has the same response in the homogenization and transmission line models. Finally, the electric field in the metamaterial and the voltage in the equivalent transmission line are linked by $V = E \times a$. Figure A1 shows the calculated steady-state envelope amplitude of the voltage at the nonlinear element as a function of $|V^{inc}|$ for port 1 and port 2 excitations, i.e. $|V_1^+| = |V^{inc}|$ and $|V_2^-| = |V^{inc}|$ (dashed curves). The curves are superimposed on the result obtained with the homogenization model developed in [17, 60] (solid curves).

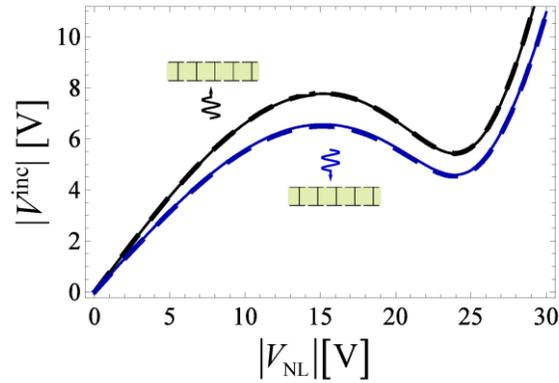

**Fig. A1.** Voltage at the nonlinear element (horizontal axis) as a function of the incident wave amplitude (vertical axis) calculated in stationary regime for $f = 14.56$ GHz. Solid curves: model of Ref. [20, 60]; Dashed curves: equivalent circuit model. The blue (black) curves are for a port 1 (port 2) excitation.

As seen, the steady-state response obtained with the transmission line model follows very closely the response predicted effective medium theory. It is relevant to mention that in Ref. [17, 60], it was shown that the effective medium model can describe very accurately the full-wave response of the system.

**Acknowledgements**

This work is supported in part by the IET under the A F Harvey Engineering Research Prize and by Fundação para Ciência e a Tecnologia (FCT) under project UID/EEA/50008/2020, and by the European Regional Development Fund (FEDER), through the Competitiveness and Internationalization Operational Programme (COMPETE 2020) of the Portugal 2020 framework, Project RETIOT, POCI-01-0145-FEDER-016432. D. E. Fernandes acknowledges support by FCT, POCH, and the cofinancing of Fundo Social Europeu under the fellowship SFRH/BPD/116525/2016.